\documentclass{aa}  
\usepackage{graphicx}
\usepackage{txfonts}
\begin{document}
\title{Spectral state transitions in low-mass X-ray binaries - the
effect of hard and soft irradiation}
\author{B.F. Liu \inst{1,} \inst{2}, F. Meyer \inst{2} and E. 
Meyer-Hofmeister \inst{2}}
\offprints{Emmi Meyer-Hofmeister; emm@mpa-garching.mpg.de}
\institute{National Astronomical Observatories/Yunnan Observatory, Chinese Academy of Sciences, P.O. Box
110, Kunming 650011, China
\and
Max-Planck-Institut f\"ur Astrophysik, Karl-
Schwarzschildstr.~1, D-85740 Garching, Germany
} 

\date{Received: / Accepted:}

\abstract{In neutron star and black hole X-ray binaries the
transitions between the two spectral states, hard and
soft, signals the change between accretion via a hot
advection-dominated flow(ADAF) and disk accretion. In a few cases the
hard/soft transition were observed during the rise to the nova
outburst, mostly only the soft/hard transition during the luminosity decrease.
Surprisingly the luminosities at the second transition is always
lower by a factor of 3 to 5. A model for this hysteresis was 
presented in a  preceding paper:
It was shown that this switch in the accretion mode at different mass 
accretion rates and therefore different luminosities is caused by the 
different amount of Compton cooling or heating of the accretion disk
corona as it is irradiated by hard or soft radiation from the
central light source, respectively.
We now give detailed results on the dependence on hardness of this 
radiation and on radiation efficiency. We further discuss the
influence of the inclination and a possible warping of the disk 
on the observed hysteresis.

}
\titlerunning {Spectral state transitions with irradiation}
\maketitle

\section{Introduction}

Of great interest in black hole physics are the properties of the matter flow 
towards a compact object. Viscous differentially rotating gas flows
determine the accretion processes around black holes, where viscous
transport of angular momentum outward allows the gas to spiral inward
towards the compact object. For a detailed discussion of the different 
possible modes of accretion see the work by Chen et al. (1995) and the 
review by Narayan et al. (1998). 

The difference in observed spectra arising from the innermost accretion flow,
can be used to study these accretion modes 
which are the same in stellar black holes or supermassive black holes
in the centers of galaxies.
Applications provide information for truncated thin disks in stellar
mass systems as e.g. the black hole binary nova Muscae 1991
(Esin et al. 1997) as well as in low luminosity active galactic
nuclei, e.g. M81 (Quataert et al. 1999).
A rich documentation is present for accretion in low-mass X-ray
binaries, containing a black hole or a neutron star primary (Chen et
al. 1997, McClintock \& Remillard 2005). The
observations display the spectra of the X-ray radiation for different
modes of accretion and correspondingly different spectral states: 
(1) a very hard spectrum (up to 100 keV) originating 
from the very energetic particles of a hot spherical
advection-dominated flow (ADAF-type) 
(2) a soft spectrum (a few keV) radiated from the much cooler 
geometrically thin accretion disk. This is a simplified picture. In
addition to the mentioned X-ray states of low and high
luminosity also an ``intermediate'' state and a ``very high'' state
can be found. McClintock \& Remillard (2005) present an overview
of the emission states of black-hole binaries where the
combined X-ray and multi-frequency spectral characteristics of the
sources are taken into account and a clearer definition of state names
is suggested.

An interesting documentation of spectral changes are the transitions 
between hard and soft state observed for
low-mass X-ray binaries, both neutron star and black hole systems 
(Tanaka \& Shibazaki 1996), and also for high-mass X-ray binaries. The first
observed change for a neutron star was found by Mitsuda et
al. (1989) for 1608-522, and for a black hole system by Ebisawa et
al. (1994), for GRS 1124-684, Nova Muscae 1991. A first modeling of
the spectra was presented by Esin et al. (1997) using the concept of an inner 
advection-dominated accretion flow. Up to now the number of
observations and the quality of spectra has very much increased, so
that these data allow us to learn from the comparison of theoretical
models with observations.

There is one special feature in the appearance of the transition of 
spectral states: considering an X-ray nova outburst light curve we see
that the hard-soft transition does not occur at the same luminosity
as the soft-hard transition. The latter happens at a
luminosity lower by a factor of about 5. In recent work
(Meyer-Hofmeister, Liu and Meyer 2005, hereafter MLM05) this
surprising hysteresis could be explained as arising from the different 
amount of Compton cooling or heating acting on the accretion disk
corona at the time of the state transition. During phases of low mass
flow in the disk, as in quiescence and the early rise to outburst, the 
inner disk region is filled with an ADAF, the radiation is very hard
and the hot photons lead to mainly Compton heating of the corona. If 
otherwise during the time around outburst maximum the mass flow rate 
in the disk is high, the thin disk reaches inward to the last stable 
orbit, the radiation is soft and the Compton effect acts as cooling of
the corona. This difference in irradiation leads to a different
vertical structure of the corona and a different amount of evaporation 
of gas from the disk.

In Sect.2 we summarize the observations for hysteresis. In Sect.3 we 
briefly describe the physics of our model for the
interaction of disk and corona and the evaporation of gas from the cool
disk.  The computation of the vertical structure of the corona 
allows to study the Compton effect on the corona in detail and to 
evaluate the dependence on the hardness of the hard radiation. In 
Sect. 4 the results of the model calculations are shown. Since the 
irradiation from the inner region affects the evaporation rate, the 
accretion rate and the efficiency of radiation enter in the results. 
We compare with observations in Sect.5. The intermediate state
between hard and soft spectral state is discussed in Sect.6, further
in the discussion a possible effect of the inclination under which we observe
a binary system in connection with flaring or warping of the
disk. In Sect.8 the conclusions follow
.

\section{Observed hysteresis in lightcurves of X-ray binaries}

Data for spectral state transitions in individual
sources were already discussed in the first paper on hysteresis (MLM05).
In Table 1 we summarize data on hysteresis  now available in the
literature. The values for the hysteresis given there are
approximate values, either derived in the work referenced
or deduced for the present paper from light curves and hardness
ratios HR2 given in the literature (state transitions taken as
occurring when HR2=(5-12keV)/(3-5 keV)= 1.5, McClintock \& Remillard
2005). The existence of a hysteresis is well documented in the 
hardness-intensity diagrams but it is difficult to deduce  
the ratio between the luminosities at the transitions.
The hardness used in these diagrams is different for
different observations and it is difficult to deduce from this
information the value of hysteresis. To deduce a reliable value for
the hysteresis is difficult also due to the fact that the time of
transition is not clearly documented, the hard/soft and even
more the soft/hard transition take some time (as discussed in following 
sections). This means the values in Table 1 only show a range. We also 
refer to the description of spectral states in the review of
McClintock \& Remillard (2005, Sect.4.3): the low/hard (LH)
and high/soft (HS) state correspond to our nomenclature hard and soft.

Historically the hysteresis effect was first pointed out by Mijamoto
et al. (1995). A first detailed analysis of hysteresis in transient 
low-mass X-ray binaries was performed by Maccarone and Coppi (2003a),
who discussed in detail the observations of Aql X-1 during the 1999
outburst, one of the best example, with two spectral transitions clearly 
documented at different count rates. For data and references of
further interesting sources see Table 1. The source 4U 1543-47 (Park
et al. 2004) was
included because it has a classical X-ray nova outburst light curve. 
McClintock \& Remillard (2005) show in their Fig. 4.2 a hard phase in
the very beginning. Since the luminosity at outburst maximum is in the 
range of the Eddington luminosity the luminosity at the hard/soft transition,
(expected at a few percent of this value) cannot be extracted from the
data. The transition soft/hard is well documented.

\begin{table*}
\caption{Observations for X-ray transients}             
\label{table:1}      
\begin {center}          
\begin{tabular}{l l l l l l}     
\hline\hline       
                      
source      &class&year&flux ratio: & keV band & references\\ 
name        &     &    &hysteresis       &          &           \\
\hline 
Aql X-1     & ns & 1999& 7  & RXTE/PCA, 1-12keV & \small{Maccarone \& Coppi (2003a})\\
            &    & 2000&10? & RXTE/PCA, 2-60keV & Maitra\& Bailyn (2004)\\
4U 1705-440 & ns & 1999& 4  & RXTE/ASM, 2-12keV & Muno et al. (2002)\\      
GX 339-4    & bh & 1998& 3.5& RXTE/ASM, 3-12keV &Zdziarski \& Gierli\'nski (2004)\\   
            &    & 2002& 10.& RXTE/ASM, 3-12keV &Zdziarski \& Gierli\'nski (2004)\\   
            &    &     & HID & RXTE/PCA, 3-21keV & Homan \& Belloni (2005)\\ 
4U 1543-475     & bh & 2002&  small? & RXTE/ASM, 2-12keV & McClintock \& Remillard(2005)\\ 
            &    &     &    & RXTE/PCA, 3-25 keV &Park et al. (2004)\\
            &    &     &    & RXTE/PCA+HEXTE, 3-200keV &Kalemci et al. (2005)\\ 
XTE J1550-564& bh& 1998& around $\ge 10^ *$&RXTE/ASM, 1-12keV&Kubota \& Done (2004) \\ 
             &   & 2000& 2$^**$ &RXTE/CA+HEXTE 2-200 & Rodriguez et al. (2003)\\
             &   & 2000& HID & RXTE/ASM, 1-12 keV & Maccarone \& Coppi (2003a)\\
XTE J1650-500& bh& 2001& 5-10 &RXTE/PCA+HEXTE, 3-150 keV& Rossi et
al. (2004)\\
             &   &     & HID  & RXTE/PCA + HEXTE, 3-200 keV& Corbel et
al. (2004)\\
             &   &     & HID  & RXTE/PCA, 3-21keV& Homan \& Belloni (2005)\\ 
GRS 1739-278& bh & 1996& 5 &RXTE/ASM, 2-12keV & McClintock \& Remillard (2005),
Borozdin et al.(1998) \\
H1743-322    &bh & 2003& HID & RXTE/PCA, 3-21keV&Homan \& Belloni (2005)\\    
GRS 1748-288 &bh &1998&HID &RXTE/PCA, 1-12keV & Maccarone \& Coppi (2003a)\\
                  &  &    &about 10   &IRS-P3/IXAE, 2-18keV  &Naik et
al. (2000), McClintock \& Remillard (2005)\\
XTE J1859+226& bh &1999&HID & RXTE/ASM, 1-12 keV& Maccarone \& Coppi (2003a)\\
             &    &    &HID & RXTE/PCA, 3-21keV&Homan \& Belloni
(2005)\\ 
XTE J2012-381& bh &1999&HID & RXTE/ASM, 1-12 keV& Maccarone \& Coppi (2003a)\\

\hline                  
\end{tabular}
\end {center}
Notes: The sources listed as black holes in the class column are
confirmed black holes or black hole candidates. The values given for
the hysteresis in the table are approximate values (see
text). $^*)$ fast rise from low/hard to very high state, no reliable
luminosity value for the hysteresis, $^**)$ transition from low state
to intermediate state. \\

HID: a hardness-intensity diagram shows the hysteresis for these
systems (different energy bands, difficult to derive a hysteresis value) 
\\
RXTE= Rossi X-ray Timing Explorer with ASM= All Sky Monitor and
 PCA= Proportional Counter Array\\
IRS-P3/IXAE= IRS -P3 satellite at the Indian X-ray Astronomy Experiment.
\end{table*}

\subsection{The hardness of observed spectra}
Since the hardness of the radiation from the inner region has an
essential influence on the evaporation process the hardness at the
time of the change from hard to soft state is of interest. But only a 
limited number of hard spectra are available for the
sources listed in Table 1.

Maccarone \& Coppi (2003b) analyzed data from the May/June
outburst 1999 of Aql X-1 and present a spectrum
taken shortly before transition to the soft state. This spectrum
ideally documents the irradiation at the state transition. We
evaluated the mean photon energy as about $h\bar{\nu}\approx$ 90 keV.
The true value might be higher depending on the poorly known spectrum at $\ge
100$ keV. In the work by Muno et al. (2002, Fig.1) hard and soft colors
for different sources are shown. The hard color of 4U 1705-440 at the
hard/soft state transition seems comparable or slightly less than that 
of Aql X-1.

Wardzi\'nski et al. (2002) analyzed hard spectra of GX
339-4 taken by the soft $\gamma$-ray OSSE detector on board CGRO
simultaneous with Ginga and RXTE observations. The energy range
reaches up to several hundred keV. The source flux at the time the
spectra were taken was about 1 percent of the Eddington luminosity, 
therefore close to a value where we expect the state transition.
(Further discussion of these spectra by Zdziarski et al., 1998 and 
Zdziarski \& Gierli\'nski, 2004). The evaluation of the spectrum from
1997 leads to $h\bar{\nu}\approx$ 100 keV. The large errors
at high energies cause quite an uncertainty in the result.

For XTE J1550-664 a hard spectrum (PCA + HEXTE data, 18-200
keV) taken 16 days before the hard/soft transition of the outburst in
2000 is shown by Rodriguez et al. (2003). We evaluate about 90 keV for 
the mean photon energy (assuming no significant contributions at higher 
energy). An averaged hard spectrum taken during the 2000 outburst 
(Arefiev et al. 2004) is less hard.

Rossi et al. (2004) present a study of the 2001/2002 outburst of the
transient source XTE J1650-500. The beginning of the outburst 
was well covered, but the hard/soft transition discussed could already
be the transition to the very high state. A spectrum taken right
before this transition (Rossi, private communication 2005) is not as
hard as the spectra from other sources discussed above.

\section{The spectral state transitions}
\subsection{The general picture}
For the analysis of the spectral state 
transitions in low-mass X-ray binaries we take the commonly accepted 
picture for the two accretion modes: (1) accretion via an optically
thick geometrically thin cool disk (cool compared to a coronal flow)
or (2) an optically thin spherically extended hot flow, ADAF-type
two-temperature solution. Usually at
larger distance from the compact object the accretion flow always is
of the first type mentioned, closer to the central object both types
are possible. The radiation is dominated by the flow in the innermost
region, that is a soft spectrum arises from disk accretion, a hard 
spectrum from the hot flow of energetic particles.
X-ray novae are good candidates to study changes between the two modes,
a hard spectrum in quiescence during mass accumulation in the disk
and a soft one after a dwarf-nova type instability had triggered an outburst
(Meyer-Hofmeister \& Meyer 1999). The importance of irradiation
had already been pointed out by de Kool \& Wrickramasinghe (1999).

Caused by the interaction of disk and corona gas evaporates from
the disk into the corona and flows in form of a hot
advection-dominated flow. Thereby the mass flow rate in the thin disk 
is diminished. The evaporation rate increases with decreasing distance 
from the compact object, but reaches a maximum at certain
distance. This maximum rate determines the switch between the
accretion modes in the inner region. Only if the mass flow in the
outer disk is larger than this value will the disk ``survive'' this
reduction in mass flow and continues inward. Otherwise
the disk becomes truncated at a certain distance depending on the mass
flow rate, and from then on all mass flows inward via the coronal/ADAF
flow. This truncation of the inner disk during phases of low mass 
flow rate has long been recognized to be an essential feature of disk 
evolution (Mineshige et al. 1998) which also appeared in numerical simulations 
(Cannizzo 1998, 2000 and Dubus et al. 2001). A recent systematic 
analysis of Done \& Gierli\'nski (2004) using all data available from 
galactic binary systems on changes of spectra and truncation radii as
a function of the accretion rate confirms this picture.

\subsection{The effect of irradiation on state transitions}
In the work by Meyer et al. (2000) the interaction of disk and corona 
was approximated by a one-zone
model incorporating the standard equations of viscous hydrodynamics 
(see also Liu et al. 2002). As became clear in the recent work on hysteresis 
(MLM05) Compton cooling and heating by radiation from 
the innermost region acting on the vertically extended corona is an 
important process. The hysteresis is mainly caused by the Compton
cooling in the soft state, but irradiation in the hard state also
plays an important role

Evaporation has three important features: (a) the rates increase towards
smaller distance $R$, (b) the rates have a maximum value at about 
several hundred Schwarzschild radii, and (c) from that distance on
inward the coupling between electrons and ions becomes poor. 
Hard irradiation leads to higher mass
evaporation rates, a higher maximal value, and therefore the spectral 
state transition at a higher luminosity than in the case of soft irradiation.
(compare Fig.1 in MLM05). The harder the hard radiation is, the higher
is the evaporation rate at maximum. Our recent computations were based
on irradiation as hard as 100keV. Since the spectra observed for
X-ray transients show quite a difference in hardness we take this into
account for the evaluation of evaporation rates.
As described in the foregoing paper (Sect. 4.3) the Compton
cooling/heating rate per unit volume taken is the sum of Compton
cooling and heating (inverse Compton and Compton effect).
\begin{equation}\label{e:compt}
 q_{\rm Comp}={4k T_e-h\bar{\nu} \over m_e
c^2}n_e\sigma_T cu,
\end{equation}
with $k$ the Boltzmann constant, $T_e$ electron temperature, 
$m_e$ electron mass, $c$ velocity of light, 
$n_e$ electron particle density, $\sigma_T$ Thomson cross section and
$u$ the energy density of the photon field. At transition the dominant
contribution comes from photons from the central source, those 
from the disk underneath can be neglected here. Also Compton cooling 
by photons of the secondary stars is generally negligible.

For the flux from the central region in the soft state we now take
a slightly different form compared to that in MLM05. We replace 
$H/R$ ($H$ scaleheight) by the term  $2\cos\delta= 2z/
(r^2+z^2)^{1/2}$,
where $\delta$ is the inclination angle under which the inner disk
appears at height $z$. The formulae then is
\begin{equation}
F={L\over 4\pi R^2}{2z\over (r^2+z^2)^{1/2}},
\end{equation}
where $L$ is the luminosity of the central
source, which is related to the central mass accretion rate $\dot M$
as $L=\eta \dot Mc^2$. The more detailed formula results in a less
strong Compton effect than the earlier used expression and therefore a
 higher maximal evaporation rate.

\section{Model calculations for hysteresis}  

In our computations we took a black hole mass $M=6M_\odot$ (the
results can be scaled for other masses and we show the results as 
measured in Eddington accretion rate 
$\dot M_{\rm{Edd}}=L_{\rm{Edd}}/0.1c^2$ with 
$L_{\rm{Edd}}=4 \pi GMc/\kappa$, $\kappa$ electron scattering opacity  
and in Schwarzschild radius $R_{\rm S} =2GM/ c^2$). The viscosity has
a strong influence on the results as shown in earlier work 
(Meyer-Hofmeister \& Meyer 2001 and Liu et al. 2002) and also pointed out by 
R\'o\.za\'nska \& Cerny (2000). We use the value $\alpha$=0.3 supported
by modeling of X-ray binary spectra (Esin et al. 1997), and also 
application to accretion disk evolution (Meyer-Hofmeister \& Meyer 1999).

\subsection{Evaporation rates  -- dependence on hardness of
irradiation and radiation efficiency}
We determined the maximal evaporation rates (the rate 
determining the state transition) in the hard state for  
30 to 120 keV mean photon energy of the radiation from the central
source. The peak value has to be consistent with the
accretion rate taken for the irradiation. For this 
consistency the efficiency $\eta$ with which radiation is produced
from accretion enters. In the recent paper we had taken the value 0.05 for the
efficiency in the hard state. 
In Fig. 1 we show the earlier results for
100 keV mean photon energy as dashed lines to illustrate the
procedure to find the peak evaporation rate for the given
efficiency $\eta=0.05$ or 0.025, marked by a filled or open dots. For the
determination of each dot a series of evaporation rates at different
distances $R$ from the center has to be evaluated to find the maximum
(and the maximum has to be consistent with respect to the radiation
produced in the central region). With this procedure we determined 
the maximal evaporation rates for different hardness and efficiency. 
The rates for $\eta=0.05$ and 0.025 are marked by filled and open dots.
Along each solid curve the efficiency decreases towards zero, without
irradiation (square).  

The cross marks the peak evaporation rate for soft irradiation when
the disk reaches inward to the last stable orbit. This is the case at 
the soft/hard transition.  
The difference between this value and the
peak evaporation rate for hard radiation yields the hysteresis value.

\begin{figure}
\centering
\includegraphics[width=8.3cm]{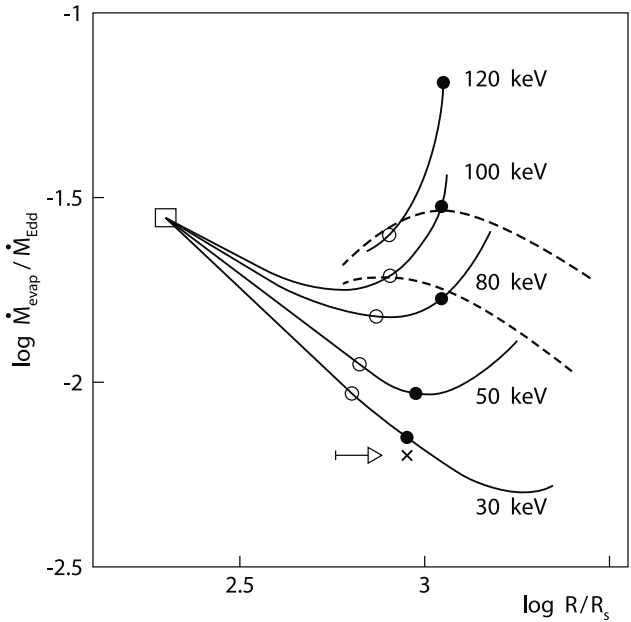}
\caption[.]{Evaporation rates at the hard/soft spectral
transition for different hardness of the radiation from the central
region. Dashed curves: evaporation rates for mean photon energy 
$h\bar{\nu}$=100 keV, upper curve radiation efficiency
$\eta=0.05$, lower curve $0.025$, dots mark maxima. Solid curves: 
Maxima of evaporation rates for $h\bar{\nu}$= 30-120 keV. Along
each curve efficiency of radiation decreases from right to left in
the diagram, values $\eta$=0.05 and 0.025 marked by filled and open
dots. The point where the curves converge (square)
corresponds to $\eta=0.$, that is no irradiation. All rates
are given in units of Eddington accretion rate, drawn at the
truncation radius $R$ (in units of Schwarzschild radius $R_{\rm S}$).
The hysteresis follows as ratio between the peak
evaporation rate in the hard state and that in
the soft state (cross).
} 
\end{figure}

\subsection{Results for spectral state transitions}
The peak evaporation rates for given hardness and efficiency of 
radiation provide the full picture of accretion rates which correspond
to the spectral transitions from hard to soft state. Taking e.g. the 
radiation efficiency as 0.05, the filled dots in Fig. 1 give the
truncation radii expected at the time of change from ADAF-type
accretion to thin disk accretion. These radii are larger than those 
without Compton effect.

\section{Comparison with observations}
\subsection{Hysteresis}
In our section on observations (Sect.2) we pointed already out
that hysteresis is clearly observed for several sources, but it is
difficult to deduce the ratio of the luminosities corresponding to the two
transition. There are several reasons. The hard/soft transition often
is not well sampled, the spectral change takes some time
(days to weeks), the exact time of transition is an interpretation of the
data. A further uncertainty enters since different energy bands are
used for the observations. The hardness-intensity diagrams presented 
by Corbel et al.(2004) and Homan \& Belloni (2005) show the effect of 
the different hardness bands used for the hardness ratios 
(4.5-7.9)/(2.5-4.6)keV and (2-200)/(3-20)keV respectively.
The observationally deduced ratios of luminosities at transition
therefore only can show a range of values. 

In Table 1 for GX 339-4 two hysteresis values are listed, 3.5 for the
1998 outburst and 10 for the 2002 outburst. The soft/hard transitions
both took place at $0.02L_{\rm{Edd}}$, a typical value, the
hard/soft transitions at $0.07$ (1998) and $0.2L_{\rm{Edd}}$
(in 2002), respectively (Zdziarski \& Gierli\'nski 2004). 
How could such a difference
arise? In our picture of state transition this always should occur at
the same luminosity and should not depend on the history as argued by 
Zdziarski \& Gierli\'nski (2004). A possible explanation could be that the
hard/soft transition at $0.2L_{\rm{Edd}}$ was already a change to the
very high state and the then earlier transition from
the low/hard to the soft state had occurred at a lower
luminosity before the observations began (comparable to the 
situation in XTE J1650, see Sect. 2.1).

Our model calculations yield a clear dependence on hardness of the 
irradiation. In Fig. 2 we show this dependence together with
luminosity ratios for some sources. As discussed in Sect. 2.1 only a
few spectra taken at the time of the hard/soft 
transition are available. The results in Fig. 2 show the wide
scatter. The hardness might be underestimated if radiation at higher 
energy bands, not included, would be present. Recent work of 
Ling (2005) reports observations of the gamma-ray emitting source 
GRO J1719-24 at energy bands up to several hundred keV. 
Similarities with other black hole X-ray transients, GRO J0422+32, and
with Cyg X-1 are mentioned and it is not clear whether  sources 
discussed here could also have X-ray emission in the gamma-ray energy region.
Then a larger hysteresis would be expected.

\begin{figure}
\centering
\includegraphics[width=8.3cm]{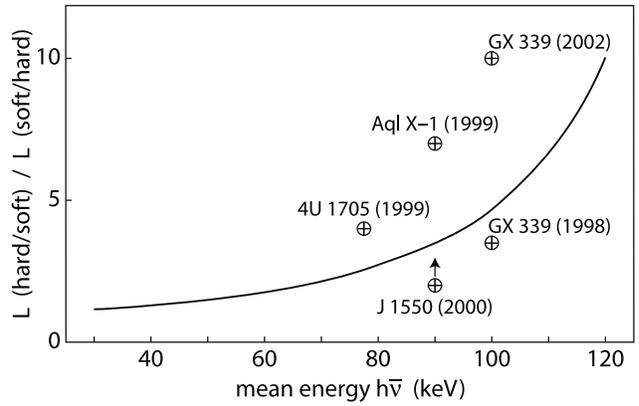}
\caption[.]{Dependence of hysteresis on mean photon energy of  
irradiation. Model (solid line) and observed values
for various sources (outburst year in parenthesis). For the
difficulty to derive these values see Sect.5. 
}
\end{figure}

\subsection{Truncation radii}
The fact that the state transitions occur at different luminosities
leads to a dichotomy in the truncation radii for this luminosity 
interval where either states can be realized: the soft state in which the
disk reaches to the last stable orbit, or, the hard state in which 
the disk is truncated according to the mass-flow rate
- radius relation. The theoretically expected truncation radii during
outburst rise and decline are shown in Fig. 2 of MLM05. Probably related to
the truncation radii are the variations in power density spectra as 
discussed by Olive et al. (2003) for 4U 1705-44.

If we compare our results for the truncation radii obtained from 
the Compton effect on the coronal structure taken into account (MLM05
and this work) we find radii larger
than expected from observations. Originally, the one-zone model for 
evaporation without including the Compton effect
yielded a disk truncation at a few hundred Schwarzschild radii when
the spectral transition occurs (Meyer et al.2000). In the paper by Yuan \&
Narayan (2004) the transition radii for several galaxies 
and galactic black hole sources are compared. Note that also
sources in the hard state were included, for which the disk truncation
at spectral state transition should be less or equal to the obtained
value. The source 
XTE J1118 is a good example, the disk edge moves inward from values of 
$10^4$ to $10^3 R_{\rm S}$ during quiescence to about 100 $R_{\rm S}$
during outburst (where the source remained in the hard state,
as can occur in case of low accretion rates (Meyer-Hofmeister
 2004)). Gilfanov \& al. (2000) estimated using  
frequency-resolved spectroscopy of Cyg X-1 that the inner radius of
the reflector, presumably an optically thick accretion disk, lies at
about 100 $R_{\rm S}$ in the hard spectral state. Based on observations 
of Nova Mus 1991 Zycki \& Done (1998) 
suggested that optically thick material within about 20-100 $R_{\rm S}$ is
generally present in the hard (low) state. The question arises whether
a left over inner disk is possible together with the ADAF, a problem
also connected with the intermediate state (see next section).

There is one parameter in our analysis which influences the 
truncation radii, the heat conduction. In forthcoming work we will 
evaluate the effect of a reduced heat conduction on the interaction of 
disk and corona. A farther effect on the coronal structure and the
location of the disk truncation might come from magnetic fields reaching
into the corona.

\section{The intermediate state}
In the foregoing sections we discussed only two spectral states, hard 
and soft. The observations document an intermediate state which can 
persist for quite some time and sometimes does not even lead to a
complete transition. Consistent 
with the explanation of the other two states this state would
correspond to an inner region with a comparable amount of mass flow 
accreting in the two forms, via a hot extended corona and via a
standard thin accretion disk. To understand how this can happen a 
closer look at the coronal evaporation model is necessary: On top of 
a standard thin accretion disk always a corona forms in interaction
with the disk, and this corona itself carries a part of the accretion
flow. Analysis of the formation of the coronal flow/ADAF
(Meyer-Hofmeister \& Meyer 2003) yielded how this coronal accretion
flow increases with decreasing distance from the black hole until
it reaches a maximum at a fixed distance between 100 and 1000
Schwarzschild radii. Inside of this distance either a pure coronal flow 
exists (if the disk became already fully evaporated further outward) 
or the thin disk continues. What happens to the coronal flow above
such a  disk continuing inward after the hot flow has passed
the distance where it reached its maximum? According to the
calculations a disk region farther in has a weaker corona and carries
less coronal flow. Therefore the part that can no longer be carried
inward must condense into the cool disk underneath (Liu et al. 2004). 

This condensation process allows interesting aspects for the
intermediate state, which we will discuss here and in forthcoming
work. Can all the former coronal flow condense so that the accretion 
in the innermost region occurs completely through the cool disk? 
This would say that in the interior we have either a pure
coronal flow in the hard state or a pure disk flow in the soft state 
with no room for the various forms of the observed intermediate state.
We suggest here a solution for this apparent contradiction between the
theoretical model and the observations. This solution relies on the 
strong temperature dependence of the collisional coupling between the 
electrons and ions in the corona. For a cool corona these two species 
are well coupled and efficiently exchange energy during the thermal 
timescale for heating and cooling of the corona. In this case
frictional heat released in the ion component is readily transfered
to the electrons and conducted to lower denser levels and radiated
away. This keeps the corona in a cool dense state as one goes to
smaller radii until it would formally disappear altogether yielding a 
perfect self consistent soft state in the interior. If however
the coronal temperature in the beginning is high, coupling between
electrons and ions is poor and the electrons can not remove the 
frictional heat of the ions which will then stay near virial
temperatures. The farther one gets inward the hotter the ions become
and the poorer is the coupling between ions and electrons. This then
will allow an intermediate state in which a hot coronal flow passes
above and below, on both sides of a regular accretion disk with not
much interaction between the two accretion streams.

Now it is very important that the regular coronal structure calculations
just yield a beginning weakening of the coupling between electrons and
ions at the distance from the black hole where the maximum coronal
flow rate occurs (compare the shape of the hatched evaporation rates 
curve in Fig.1). If the accretion rate is higher than this maximum
the disk does not fully evaporate and the coronal flow condenses 
effectively into the cool disk further in. If however the
accretion rate drops to a value lower than the maximal possible coronal
flow rate the disk evaporates over some distance outside of
the radius of the evaporation maximum and a gap is formed. From the 
outer border of the gap on through the whole gap region the 
accretion continues inward in form of an ADAF. ADAF temperatures are
near  virial and are higher than those of coronae above a 
cool accretion disk at the same distance from the central black hole. 
Thus where the ADAF reaches the inner boundary of the gap i.e. the  
inner disk, it has a rather high temperature. This temperature is
higher the farther in the gap extends because the 
virial temperatures increases with $1/R$ as the radius $R$ decreases.
Correspondingly poor becomes the coupling between the electrons and ions 
and the flow is therefore only partially capable to cool off
and to condense back into the disk.
Though this partial condensation will feed the disk and keep it
stretching all the way inward another part of the accretion flow will 
stay in the hot near virial flow above and below the disk and on
arriving in the interior will give the hard spectral component to this 
intermediate state, the soft component being provided by the remaining 
disk flow.

If the accretion rate is quite close to the critical value (that is
the maximal evaporation rate) or above it 
the hard power law tail will only constitute a small fraction of the 
spectrum and with the fluctuating accretion rate vary significantly. 
This would then account for the observed hard power law tail in the soft
state.

\section{Discussion}

A number of factors can complicate a direct comparison between 
theoretical model and observations.

In the soft state the luminosity is radiated from a thin innermost
accretion disk. When this disk lies in the equatorial plane the
observer sees it foreshortened. The "observed" luminosity is then
inclination dependent and must be corrected for this projection
in order to obtain the true luminosity at the soft/hard transition.

Further, should the inner disk be tilted out of the equatorial
plane and precess, an additional and time varying aspect of the
accretion disk enters and would simulate an apparent variation of
the transition luminosity even when the true transition luminosity
remains the same. Tilted and precessing accretion disks might be
indicated by periodic shifts in color-intensity plots of low mass
X-ray binaries.

Recently Narayan and McClintock (2005) investigated the effects of
the inclination angles of black hole X-ray binary systems on their
observational properties. Within their sample of 20 sources
they found none with an inclination angle larger than $75^\circ$ and
suggested that this absence of eclipsing sources is due to a flaring of
the accretion disk by about $15^\circ$, so that the disk permanently
occults the X-ray source for the observer for all inclination angles
that would allow us to see eclipses. Noisy light curves of sources with
inclination between 70 and $75^\circ$ support this picture and
indicate the presence of partial and time varying coverage of the
central source. Such a partial occultation could affect the
determination of the luminosities at both hard/soft and soft/hard
transitions.

A perhaps more remote possibility which would however have a direct
impact on the coronal evaporation mechanism itself and thus on the
hysteresis is a warping of the inner accretion disk. This would change
the aspect under which the corona sees the irradiating innermost disk
surface and thereby change the strength of the irradiation.

\section{Conclusions}

As one follows  the light curve of an X-ray nova from
early rise to late decline, observations show a profound
change in the source spectrum from a hard to a soft one
on the rise and back from a soft to hard one on the
decline. These changes occur at characteristic luminosities
but remarkably, they are not the same for the two
transitions. This hysteresis in the light
curve could be understood as arising from the different type
of irradiation coming from the innermost region, which is
hard at the hard/soft transition but soft at the soft/hard
transition: This difference leads to a different Compton
cooling or heating of the coronal layers and results in a
different coronal density and mass flow. The latter determines
whether a disk can be truncated (hard state) or continues
all the way to the black hole (soft state)(MLM05).

Now we have carried out further analysis and show how
this hysteresis depends on the hardness of the irradiation.
Unfortunately, only a few spectra taken at the moment of
state transition are available. Also the contributions at the
important high energies are uncertain. Thus the
observational basis for a comparison is still fairly sparse.
In addition theoretical uncertainties arising from the rough
"one-zone" modeling and the choice of its parameters
together with the assumed planar inner disk geometry will
enter into comparisons between observations and theoretical
models.

Thus to verify in detail the theoretically well understood
dependence on the hardness of the hard irradiation in the
comparison with the data is difficult.
It is difficult to determine when and at which luminosity
the transition actually happened. Different observations are taken in different
energy bands. Since the change from one state to the
other takes some time this clearly also demands a more
detailed understanding of the intermediate state between
the hard and the soft state. In our first discussion here
we find a remarkable coincidence of the distance where
the coronal flow rate reaches its maximum with that of
the least thermal coupling between electrons and ions
which could perhaps explain the features of an
intermediate state.

In conclusion one might take the explanation of the
hysteresis as a further strong support for the coronal
evaporation model. It is surprising how this very
simplified model appears to catch essential features of
the accretion flows around compact objects. More
observations of spectral state transitions in different
systems, broad spectral energy coverage and refinement
of the theoretical modeling in the future might prove
very fruitful.  

\begin {acknowledgements} 

One of the authors, BFL, thanks the Alexander-von-Humboldt Foundation 
for the award of a research fellowship during which this investigation started.
We would like to thank Sabrina Rossi and collaborators for 
providing a spectrum of XTE J1650 at the time of transition during
outburst rise.
\end {acknowledgements}

\end{document}